\documentclass[aps,prl,reprint,floatfix,superscriptaddress]{revtex4-1}
\usepackage{amsmath,amssymb}
\usepackage{float}
\usepackage{graphicx}
\usepackage[utf8]{inputenc}
\usepackage{hyperref}
\usepackage{color}
\hypersetup{colorlinks=true,citecolor={blue},linkcolor={blue},urlcolor={blue}}
\usepackage{natbib}
\graphicspath{{images/}}

\hypersetup{colorlinks=true,citecolor={blue},linkcolor={blue},urlcolor={blue}}
\graphicspath{{images/}}

\begin{document}

\title{Dynamics of the Energy Relaxation in a Parabolic Quantum Well Laser}
\author{A. V. Trifonov}
\email[correspondence address: ]{arthur.trifonov@gmail.com}
\affiliation{Spin Optics Laboratory, St. Petersburg State University, 1, Ulianovskaya, St. Petersburg, 198504, Russia
}
\author{E. D. Cherotchenko}
\email[correspondence address: ]{E.Cherotchenko@soton.ac.uk}

\author{J. L. Carthy}
\affiliation{ School of Physics and Astronomy, University of Southampton, Southampton, SO17 1BJ, United Kingdom
}
\author{I. V. Ignatiev}
\affiliation{Spin Optics Laboratory, St. Petersburg State University, 1, Ulianovskaya, St. Petersburg, 198504, Russia
}
\author{A. Tzimis}
\affiliation{ Department of Materials Science and Technology, University of Crete,
71003 Heraklion, Crete, Greece
}
\affiliation{FORTH-IESL, PO Box 1385, 71110 Heraklion, Crete, Greece
}
\author{S. Tsintzos}
\affiliation{ Department of Materials Science and Technology, University of Crete,
71003 Heraklion, Crete, Greece
}
\affiliation{FORTH-IESL, PO Box 1385, 71110 Heraklion, Crete, Greece
}
\author{Z. Hatzopoulos}
\affiliation{ Department of Materials Science and Technology, University of Crete,
71003 Heraklion, Crete, Greece
}
\affiliation{FORTH-IESL, PO Box 1385, 71110 Heraklion, Crete, Greece
}
\author{P. G. Savvidis}
\affiliation{ Department of Materials Science and Technology, University of Crete,
71003 Heraklion, Crete, Greece
}
\affiliation{FORTH-IESL, PO Box 1385, 71110 Heraklion, Crete, Greece
}
\author{A.V. Kavokin}
\affiliation{ School of Physics and Astronomy, University of Southampton, Southampton, SO17 1BJ, United Kingdom
}
\affiliation{Spin Optics Laboratory, St. Petersburg State University, 1, Ulianovskaya, St. Petersburg, 198504, Russia
}

\date{\today }

\begin{abstract}
We explore two parabolic quantum well (PQW) samples, with and without Bragg mirrors, in order to optimise the building blocks of a Bosonic Cascade Laser. The photoluminescence spectra of a PQW microcavity sample is compared against that of a conventional microcavity with embedded quantum wells (QWs) to demonstrate that the weak coupling lasing in a PQW sample can be achieved. The relaxation dynamics in a conventional QW microcavity and in the PQW microcavity was studied by a non-resonant pump-pump excitation method. Strong difference in the relaxation characteristics between the two samples was found. The semi-classical Boltzmann equations were adapted to reproduce the evolution of excitonic populations within the PQW as a function of the pump power and the output intensity evolution as a function of the pump-pump pulse delay. Fitting the PQW data confirms the anticipated cascade relaxation, paving the way for such a system to produce terahertz radiation.
\end{abstract}

\maketitle

%Abstract complete,insert pacs.

%PRL and PRD papers have to have PACS (Phsyics and Astronomy Classification Scheme) numbers. Please see {\tt http://www.aip.org/pacs/} for the numbers relevant to your paper. A set of standard references can be found at the end of this example paper.
%\maketitle must follow title, authors, abstract, \pacs, and \keywords
%------------------------------------------------------------------------------------------------
%%  INTRODUCTION  %%

\section{Introduction}

Recently research into coherent light sources based on bosonic systems (known as bosers, or bosonic lasers) has seen a rapid increase \cite{kavokin_polaritons_2013}. In contrast to conventional lasers based on the
phenomenon of stimulated emission, bosonic lasers are based on stimulated
relaxation of bosons and the formation of an exciton-polariton condensate \cite{savona_fifteen_2008}.
This stimulated relaxation is triggered by the final state occupation of an
energy level within a system, and serves as the principal tool for building
up of a polariton population in a given energy state \cite{weihs_excitonpolariton_2003}. The coherence of boser
radiation is the result of spontaneous emission of photons by the condensate
after its occupation exceeds unity~\cite{kasprzak_boseeinstein_2006} making
such a system ideal for a low-threshold lasing device.

In this paper we build upon the idea of a bosonic cascade laser (BCL)
introduced by Liew et al.~\cite{liew_proposal_2013} that is capable of
emitting terahertz (THz) radiation, a technologically under-developed
section of the electromagnetic spectrum \cite{tonouchi_cutting-edge_2007}. The BCL uses a cascade mechanism similar to that of the quantum cascade
laser (QCL)~\cite{faist_quantum_1994, bogdanov_mode_2011-1} in order to
generate radiation. Unlike the QCL, which uses multiple adjacent quantum wells (QWs)~\cite{kazarino.rf_possibility_1971} as the cascade ladder, the
BCL cascade~\cite{kaliteevski_double_2012} is formed by equidistant
excitonic levels in a single parabolic quantum well (PQW)~\cite{ulrich_temperature_1999,studer_gate-controlled_2009,geiser_ultrastrong_2012,duque_electron_2013-1,phuc_nonlinear_2015,yan_optical_2015}. Although intersubband polariton QCL lasers have been proposed \cite{colombelli_quantum_2005, de_liberato_stimulated_2009-1}, these rely upon the need for population inversion between adjacent subbands, analogous to the QCL. In a BCL, however, the amplification is due to the bosonic stimulation of radiative transitions
between levels in a cascade. Both the QCL, and a range of other proposed microcavity systems are capable of generating THz \cite{kavokin_vertical_2012-1, del_valle_terahertz_2011, savenko_nonlinear_2011-1,kavokin_stimulated_2010}, but the BCL uniquely offers increased amplification created by the final polariton state stimulation within the confines of \textit{one} PQW.

We report on the realisation of such PQW heterostructures, with and without
external microcavities (MCs), and explore the optical characteristics of the
PQWs compared to a MC sample with conventional rectangular quantum wells (QWs). We
experimentally demonstrate that PQWs in a MC are capable of acting as
non-resonantly excited polariton lasers, as well as being able
to emit light via the bosonic cascade mechanism. We find unusual
pump-power dependencies of the photoluminescence (PL) in the PQW sample
without MC, which we believe to be due specifically to the bosonic cascade
relaxation mechanism. We also investigate the relaxation dynamics of
excitons in MCs with parabolic and rectangular QWs. Using a pump-pump
method, excitons are seen to relax in the PQW much faster than in a MC with
a rectangular QW. We believe the accelerated relaxation in PQW to be an
indication of stimulated relaxation in a bosonic cascade and find that the
experimental results are in agreement with the BCL model of ref.~\cite{liew_proposal_2013}. 	
%------------------------------------------------------------------------------------------------
%% EXPERIMENTAL SET UP AND CHARACTERISATION %%

\section{Samples and Experimental Setup}

%% SAMPLES %%
A PQW sample without and with (denoted S1 and S2 respectively) Bragg
reflectors (DBRs) have been studied, and their relaxation and excitation
characteristics have been compared to a planar microcavity sample with
square QWs (S3). All samples were fabricated with molecular
beam epitaxy. S1 contains an InGaAs/GaAs PQW of width $\approx 50$~nm at the
top of the potential well, and the parabolic profile was achieved by
altering the indium concentration during the growth process from 2\% at the InGaAs/GaAs interface to 6\% in the
middle of QW (see inset in \mbox{Fig.\,~\ref{fig:1}}). Sample S2 was
fabricated similarly with an Al$_{x}$Ga$_{1-x}$As/Al$_{0.15}$Ga$_{0.85}$As
QW of $\approx 50$ nm width, where the parabolic profile was
achieved by altering the concentration of aluminium along the $z$-axis
of the sample from 5\% in the middle of the QW to 12\% near the interface.
The microcavity was formed with two DBRs, each with 17 and 22 Al$_{0.15}$Ga$%
_{0.85}$As/AlAs paired layers. The Q-factor of the microcavity is approximately
2,000 and the PQW was placed in the middle of $3\lambda /2$ intracavity
spacing.

Finally a $5\lambda/2$ planar GaAs cavity, sample S3, consisting of 32 and 35 Al$_{0.15}$Ga$_{0.85}$As/AlAs DBR pairs and 12 rectangular
QWs was studied. The Q-factor of this microcavity is approximately 16,000.

%% SET UP %%
The samples were mounted in a close-cycle cryostat to reach a temperature of
roughly \mbox{5\,K}. Samples S2 and S3 were excited non-resonantly above the microcavity stop-band by femtosecond
pulses from a Ti:Sa laser. The
laser spot size was approximately \mbox{30\,$\mu$m}. S1 was
exited with a CW laser resonantly tuned to the exciton resonance in the
barrier layers (to the $12^{th}$ quantum confined excitonic state, see Fig.~%
\ref{fig:1}). Such pump conditions allowed us to create excitons rather than
electron-hole pairs. To study the exciton relaxation dynamics of S2 and S3
we used a pump-pump technique, whereby two pump pulses separated by a variable
delay are used to excite a sample non-resonantly with great temporal
resolution giving the time-integrated intensity of the PL as a function of
the delay. All PL spectra and $k$-space images were recorded by a 0.55 m
imaging spectrometer equipped with a CCD.

%------------------------------------------------------------------------------------------------
%%%%%EXPERIMENTAL RESULTS%%%%%%%%%%%%

\section{PQW without microcavity.}

\begin{figure}[h]
\centering
\includegraphics[width=0.42\textwidth]{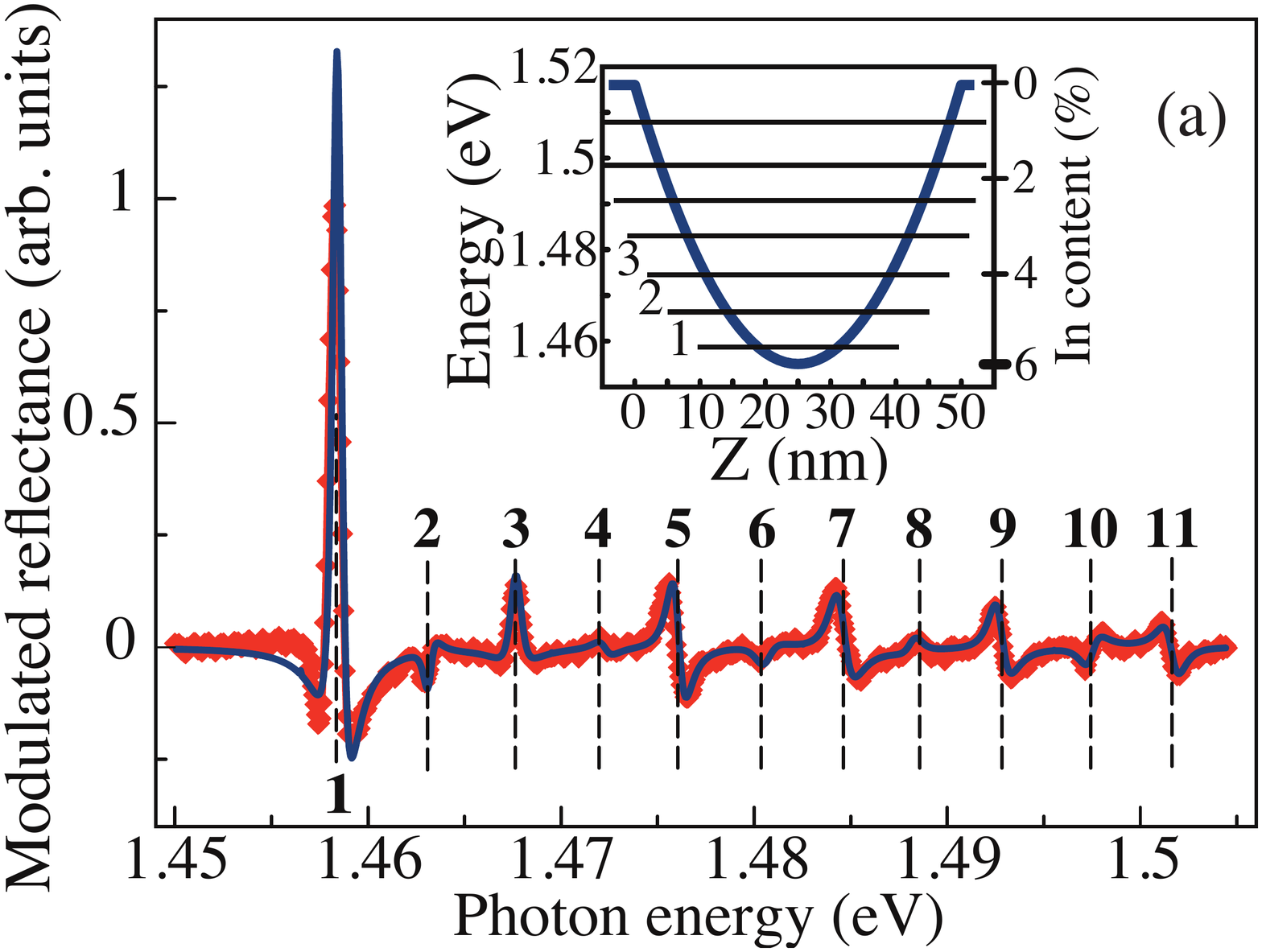}  \includegraphics[width=0.4%
\textwidth]{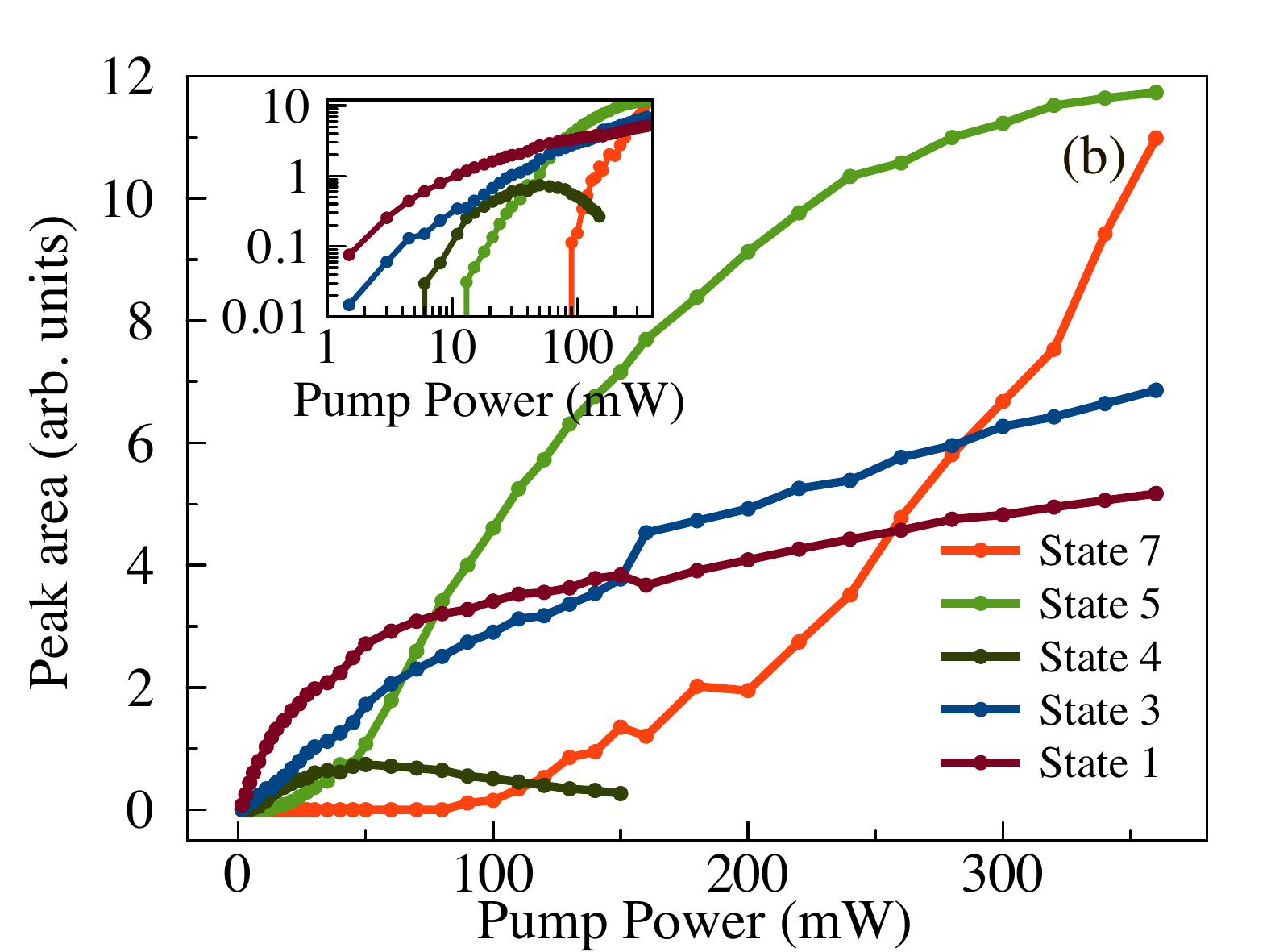}
\par
%\rule{35em}{0.5pt}
\caption[PL for pqw]{(color online) (a) $\protect\lambda$-modulated reflectivity spectrum
of sample S1 containing parabolic QW without microcavity (red curve) and
modelled spectrum (black curve). Vertical dashed lines mark equidistant
quantum confined excitonic states in the PQW. Inset: the potential
profile for excitons (left axis) and distribution of indium content across
the QW layer (right axis) are shown. (b) Pump power dependencies of integral
PL from different quantum confined excitonic states. The pump wavelength was
tuned to the exciton resonance in the barrier layer. The integral PL for
each transition was obtained by deconvolution of the PL spectra into a set
of Lorentzians.The inset presents the same curves plotted in logarithmic scale to show the low power region.}
\label{fig:1}
\end{figure}

In order to realise a BCL and confirm its relaxation mechanism, we first
characterised the bare PQW sample, S1, to verify that equidistant exciton states had been achieved. Therefore S1 was assessed
using sensitive $\lambda$-modulated reflection~\cite{CardonaModSpectr} and
PL spectroscopy. In \mbox{Fig.\,~\ref{fig:1}~(a)} the modulated reflectance
spectrum is presented (red curve) and subsequently fitted (solid black line)
and up to 11 distinct excitonic states can be resolved. The energy spacing
between the neighboring resonances is about \mbox{6\, meV}, or \mbox{1.45\, THz}. The inset in
Fig.~\ref{fig:1}~(a) shows the potential profile for excitons in PQW (blue
line) and the positions of equidistant quantum confined excitonic states
(horizontal black lines), which creates the bosonic cascade ladder.

%%% REFLECTION SPECTRUM %%
We performed a simple analysis of the reflectivity spectrum generalizing the
theory developed in refs.~\cite{Ivchenko-book, Trifonov2015} for the case
of several exciton quantum confined states to fit the modulated reflection
data. The modulation technique was used to reduce a noise and to stretch weak features connected to the excited quantum confined excitonic states. Following the non-local dielectric response theory,\cite{Ivchenko-book}
the amplitude reflection coefficient for a QW with several exciton
resonances can be written in the form: 
\begin{equation}
r_{QW}=\sum\limits_{N=1}^{N_{max}}\frac{i(-1)^{N-1}\Gamma _{0N}e^{i\varphi
_{N}}}{\omega _{0N}-\omega -i(\Gamma _{0N}+\Gamma _{N})}.  \label{IvchenkoEq}
\end{equation}%
Here $\omega _{0N}$ is the resonance frequency, $\Gamma _{0N}$ and $\Gamma
_{N}$ are the radiative and nonradiative damping rates for a system of $N$
levels. The phase $\varphi _{N}$ in this equation takes into account a
possible asymmetry of the QW potential. Reflectivity $R(\omega )$, from the
structure with a top barrier layer of thickness $L_{b}$ and a QW layer of
thickness $L_{QW}$, is calculated using the transfer-matrix approach: 
\begin{equation}
R(\omega )=\left\vert \frac{r_{01}+r_{QW}e^{2i\phi }}{1+r_{01}r_{QW}e^{2i%
\phi }}\right\vert ^{2},  \label{Reflectance}
\end{equation}%
where $r_{01}$ is the amplitude reflection coefficient from the sample
surface. The phase is $\varphi =K(L_{b}+L_{QW}/2)$, where $K$ is the photon wave
vector in the heterostructure. The calculated derivative reflectivity
spectrum is shown in Fig.~\ref{fig:1} (black curve). 

 The peak integrated PL spectra of the sample S1 at different excitation powers has been
measured (see Fig~\ref{fig:1}~(b)). We have found that the power increase gives
rise to the increase of PL intensity from the lowest exciton state followed
by its saturation, contrary to the model set out in Ref.\cite{liew_proposal_2013} where the highest level is seen to populate first, and the lowest level establishing a population last. Simultaneously, the intensity of the PL from the excited
exciton states increases super-linearly with pump power and then also
saturates. The similar behavior of PL is observed for exciton states under
further increase of the pump power. The full set of the PL data consisting
of about 500 spectra was analyzed by deconvolution of each spectrum into a
set of Lorentzian resonances, corresponding to different exciton transitons. 
We have found that such deconvolution fits the experimentally
observed spectra, if the wavelength of excitation coincides with one of the
exciton resonances in PQW or with the exciton resonance in barrier layers.
In this case, the pump directly creates excitons rather than uncoupled
electron-hole pairs, the relaxation of which differs from exciton relaxation. If
electron-hole pairs are created by the non-resonant excitation, a broad
structureless background appears in the PL spectra.

We should stress that the observed behavior of the time integrated PL
intensities of resonant exciton peaks is a characteristic of PQWs. This is a
clear indication that excitons created by resonant excitation relax via
cascade between neighboring energy levels. We have studied by the same
technique a reference rectangular QW of thickness of about 90 nm and found
that no new exciton lines appear in the PL spectra with the pump power
increase.

\section{PQW in a microcavity.}

The ability of S2 to act as a polariton laser~\cite%
{weihs_excitonpolariton_2003,kavokin_semiconductor_2003-1} is a non-trivial
question. Heterostructures acting as polaritonic lasers usually contain
multiple thin QWs (of roughly 10 nm in width) to increase the oscillator
strength of the excitonic transition in order to establish strong coupling. S2, however, contains just one PQW of about 50 nm width. The larger QW thickness and stronger overlap
of the electron and hole wave functions in the PQW provides a
sufficiently large exciton-photon coupling to make the strong coupling
regime and polariton lasing possible \cite{bajoni_polariton_2008,butte_transition_2002}.

\begin{figure}[h]
\centering
\includegraphics[width=0.4\textwidth]{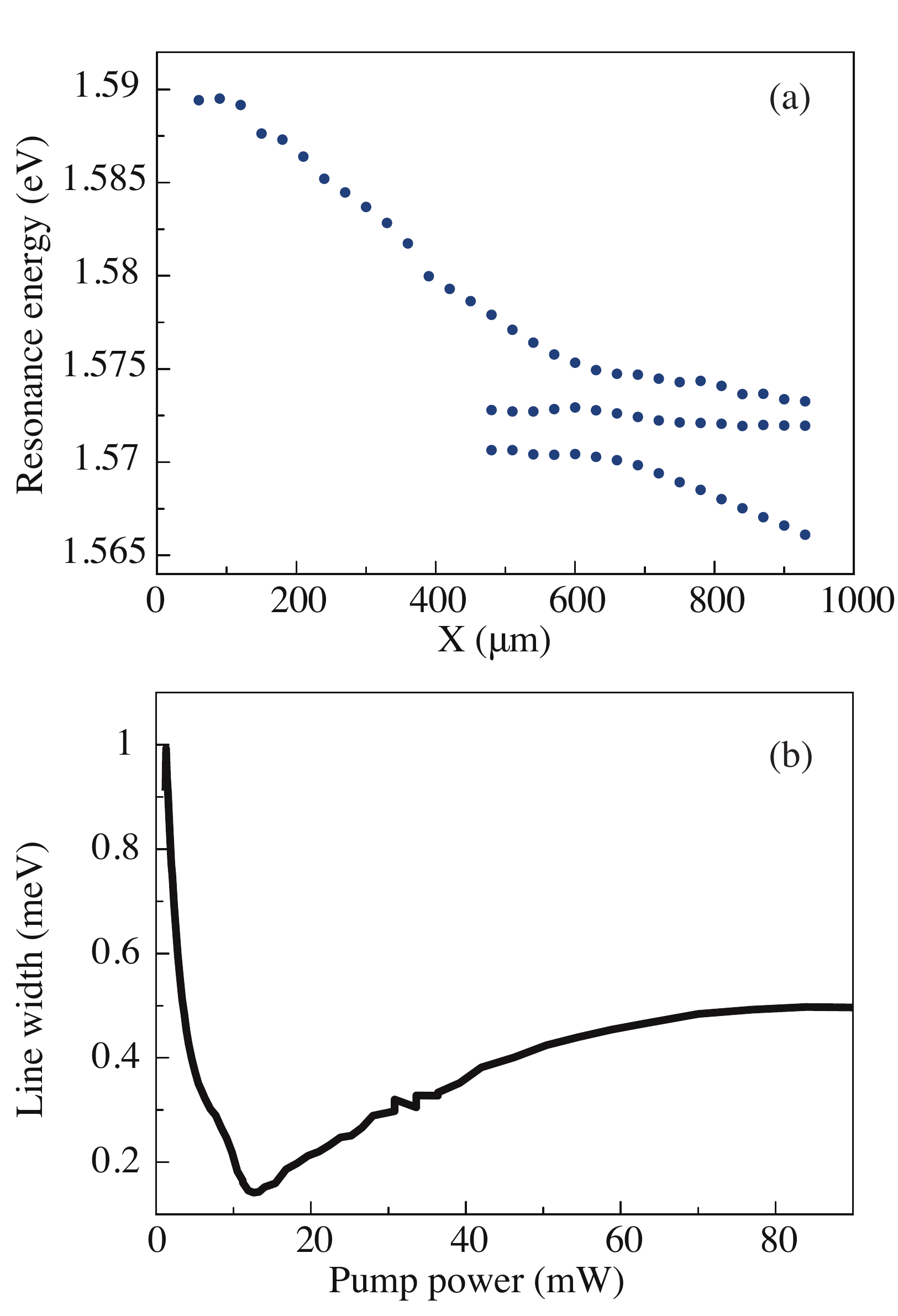}  %\rule{35em}{0.5pt}
\caption[PL for pqw]{(color online) (a) The energy position of features in reflectance
spectra of sample S2 as a function of the laser spot position on the sample. The figure
demonstrates anti-crossing, a classic signature of the presence of
exciton-polaritons. (b) The PL linewidth dependence on the excitation power
for sample S2 is seen to narrow significantly at the lasing threshold due to
increased coherence of the exciton-polaritons within the sample's polariton
trap.}
\label{fig:2}
\end{figure}

\begin{figure}[h]
\includegraphics[width=0.4\textwidth]{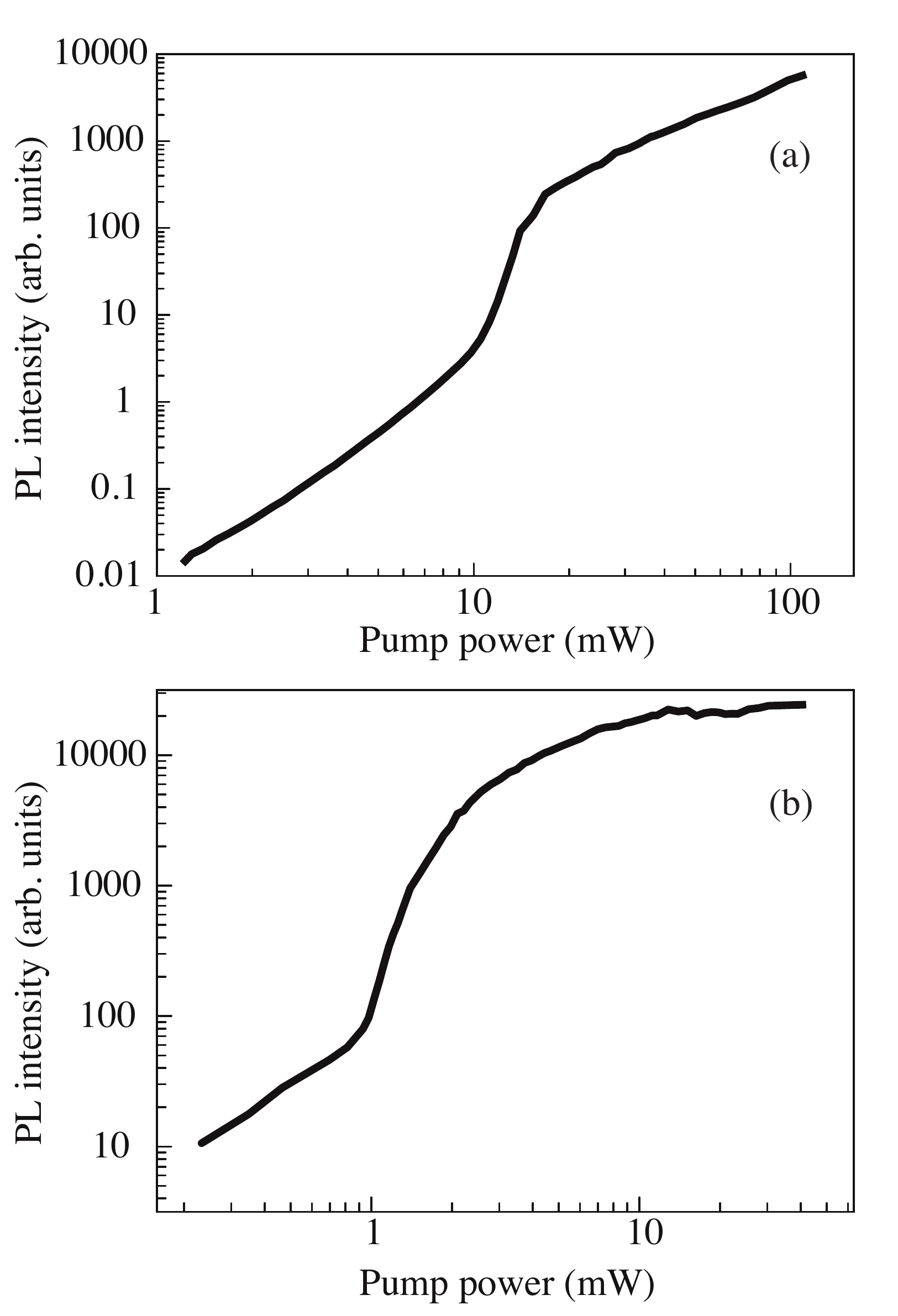}
\caption[PL for pqw]{(color online) The PL intensity measured as a function of the pulsed
excitation power for (a) PQW (sample S2) and (b) sample S3 with multiple
rectangular QWs in MC. }
\label{fig:3}\centering
  %\rule{35em}{0.5pt}
\end{figure}

Fig.~\ref{fig:2}~(a) shows the dependencies of polariton mode energies on the
laser spot position on the sample S2. One can see that the detuning between
the exciton and photon resonances is dependent on the spot position. The
anticrossing of polariton modes is clear evidence of the strong
coupling regime. In the anticrossing range the reflectivity spectrum
exhibits three distinct minima, these can be attributed to the coupling of
the heavy-hole and light-hole excitons\cite{mathieu_differential_1987-1} to the cavity mode. The Rabi
splitting of the relating polariton states is about 6 meV.

The pump-power dependence of PL intensities for samples S2 and S3 are shown
in Fig.~\ref{fig:3}. For both samples the threshold-like increase of the
intensity is clearly observed. For the sample S3 the threshold manifests the
polariton lasing regime. For the sample S2, the identification of the
threshold is questionable without additional experiments. Note, that the PL
intensity rises exponentially with pump power below the threshold. At the
pulsed excitation this may be an indication of the switching of the system
to the lasing regime within a limited time- window, which becomes larger as
the pump power increase.We consider it as an indication that the stimulated
relaxation occurs in sample S2. 

%%%%%%%%%%%%%%%%%%%%%%%%

\section{Pump-pump experiments}

The pump-pump method we employ allowed one to highlight relaxation processes
in the PQW within the MC that may be hidden for studies by conventional time
resolved spectroscopy methods due to the reflection from the DBRs\cite{damen_dynamics_1990}. The important property of the stimulated cascade relaxation is its strong dependence on population of the lower-lying exciton state~\cite{liew_proposal_2013}. In the pump-pump method, the first pulse creates some initial density of excitons and the delayed second pump pulse creates
additional excitons at the pumped level. Relaxation of these excitons
strongly depends on the population of lower energy excitonic levels created
by the first pump pulse. If this population is large enough, the stimulated
relaxation is triggered and accelerated. This acceleration should result in
the nonlinear increase of the total PL signal excited by both pump pulses in
the case of competing radiative and non-radiative channels of polariton
recombination. The PL intensity should depend on the delay between
two pulses; no nonlinear PL increase should occur at very
large delays, where the excitons created by the first pulse relax and
recombine before the second pulse arrives.
\begin{figure}[h]
\begin{minipage}[h]{0.49\linewidth}
\center{\includegraphics[width=1\linewidth]{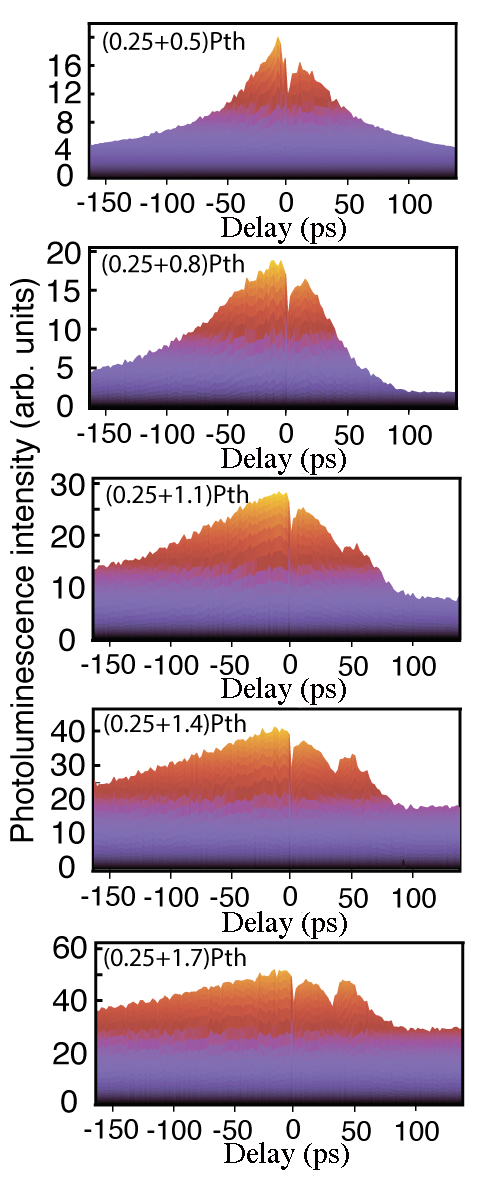} \\ (a)}
\end{minipage}
\hfill 
\begin{minipage}[h]{0.49\linewidth}
\center{\includegraphics[width=1\linewidth]{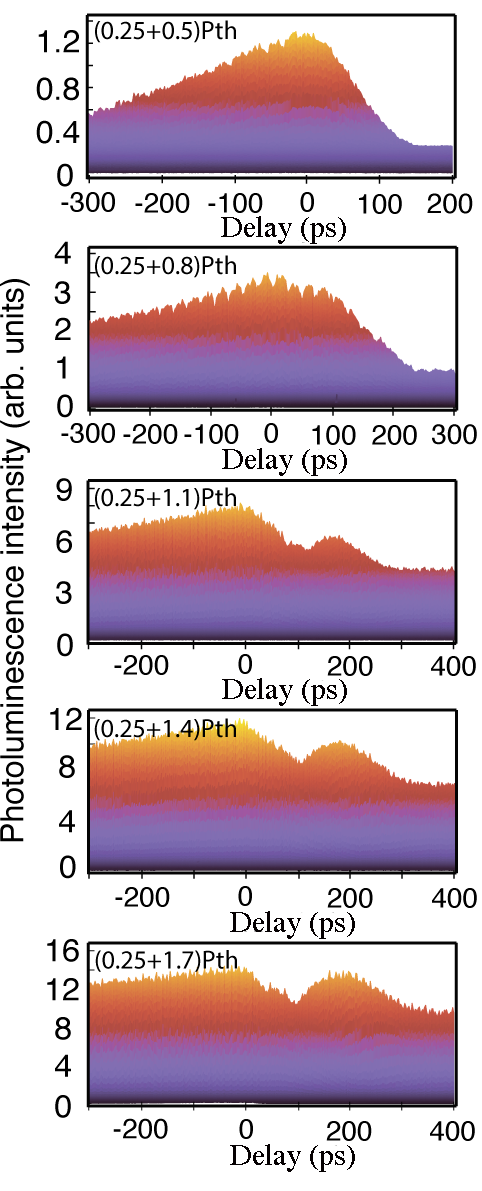} \\ (b)}
\end{minipage}
\caption{(color online) Photoluminescence intensity plotted as a function of a delay
between two fs pulses a)PQW (S2) and b) rectangular QW (S3).}
\label{fig:5}
\end{figure}
Thus the samples S2 and S3 were studied by the pump-pump method. Samples in
the cryostat were pumped by two femtosecond pulses. The total time
integrated PL intensity was measured as function of delay between two pulses.

Fig.~\ref{fig:5} shows the delay dependencies of PL intensity measured at
different excitation powers of both pulses for samples with PQW (S2) and
multiple rectangular QWs (S3). The pumped powers are chosen close to the
threshold of lasing. There are several peculiarities in these dependencies.
Firstly, there is a strong increase of the PL at a relatively small delay.
At the same time, the range of these delays is considerably smaller for PQW
(150 $ps$) than that for rectangular QWs (several hundred $ps$). These means
that the relaxations processes in the PQW is considerably faster than in the
rectangular QWs.

Secondly, additional features in these dependencies are observed. For PQW, a
relatively narrow dip at zero delay is clearly seen. No such dip is observed
for rectangular QWs. This is an indication of significant difference in
relaxation processes in these two samples. 

Thirdly, some asymmetry of the dependencies for positive and negative delays
is observed in both samples. It is caused by the difference in pump powers
of first and second pulses. If the weak pulse comes first (negative delay),
the exciton population is relatively small, consequently the bosonic
stimulation is weak, and the relaxation of excitons created by the weak
pulse is relatively slow. When the strong pulse comes first (positive
delay), the stimulated relaxation is accelerated compared to the case of a
weak pulse coming first.

Finally, an additional peak of PL intensity is observed at the positive
delay, if the power of the strong pulse is beyond the threshold.

To understand the second peak at the positive delays one should discuss the
time dependence of PL intensity. In ref.~\cite{Bloch}, the PL kinetics has been studied at different exciting powers for both below and above the threshold of polariton lasing, $P_{th}$. It was found that, when the power $P < P_{th}$, the PL intensity slowly rises and reaches its maximum at $t_1 \approx 100$~ps. At later
times PL intensity slowly decreases with characteristic decay time $t_2 \approx 400$~ps. When the pump power exceeds the threshold, a strong pulse of polariton laser emission appears at time $t_1$ with the 10-20 $ps$ pulse duration. Such temporal behavior of PL intensity allows us to assume the following origin of second peak in the pump dependencies shown in Fig.~4. When the sample is pumped by two pulses
and the first pulse power is above the threshold two maxima of the polariton
laser emission may be seen. The reason is that the number of excitons remained
after the first pulse of polariton lasing peak and of excitons created by
the second pump pulse is sufficient for the formation of the second peak of
polariton laser emission. These effects with polariton lasing appearing
twice are expected to be present in the sample S2 with PQW in MC as well as
in the sample S3 with rectangular QW in MC. But the time delay and the width
of the second peak strongly depends on the relaxation dynamics. Thus in the
sample S2 the relaxation is faster than in the sample with a rectangular QW
 in MC.\\
 
\section{Modelling}
Exciton relaxation and dynamics in GaAs MCs has been extensively studied in conventional QWs~\cite{schultheis_ultrafast_1986,heller_exciton_1995, khitrova_nonlinear_1999,bongiovanni_coherent_1997}, particularly in what concerns phonon mediated relaxation. 
In order to analyze the experimental data obtained for the PQW, we use with the rate
equations introduced by Liew \textit{et al}.~\cite{liew_proposal_2013}, using the  Boltzman kinetic theory of relaxation~\cite{deng_exciton-polariton_2010}. We consider $m$ distinct excitonic levels in a PQW. 
The dynamics of population of each of the levels can be described by the following system of rate equations:  
\begin{align}
&\frac{dN_{m}}{dt}=-\frac{N_{m}}{\tau _{u}}%
-\sum_{i=1}^{m-1}W_{i}N_{m}(N_{m-i}+1) \nonumber \\
 &  + \frac{\alpha}{2}N_{m-1}^{2}(N_m + 1)(N_{m-2} + 1) - \frac{\alpha}{2}N_mN_{m-2}(N_{m-1} + 1) \label{eq:1}
\end{align}%
\begin{align}
&\frac{dN_{k}}{dt}=P_{k}^{(1)}(0) + P_{k}^{(2)}(\tau_{delay})-\frac{N_{k}}{\tau _{p}}+  \notag  \label{eq:2} \\
&	+\sum_{i=1}^{m-k}W_{i}(N_{k+i}(N_{k}+1))-\sum_{i=1}^{k-1}W_{i}N_{k}(N_{k-i}+1)\nonumber \\
& 	- \alpha{N_k}^2(N_{k+1} + 1)(N_{k-1}+1) +\alpha{N_{k+1}}N_{k-1}(N_k+1)^2\nonumber \\
&  	+ \frac{\alpha}{2}{N_{k+1}^2(N_{k} + 1})(N_{k+2} + 1) - \frac{\alpha}{2}(N_{k+1} + 1)^2N_{k}N_{k+2} \nonumber \\
& -\frac{\alpha}{2}{N_{k-1}^2(N_{k} + 1})(N_{k-2} + 1) + \frac{\alpha}{2}N_{k-2}N_k^2({N_{k-1}} + 1)^2 \nonumber \\ 
 &k=2..m-1,
\end{align}%
\begin{align}
&\frac{dN_{1}}{dt}=-\frac{N_{1}}{\tau _{g}}+\sum_{i=1}^{m-1}W_{i}{N_{i+1}}(N_{1}+1) \nonumber \\
&	+ \frac{\alpha}{2}N_2^2(N_1 + 1)(N_3 + 1) - \frac{\alpha}{2}N_1N_3(N_2 + 1)^2 \notag \\
\label{eq:3}
\end{align}%

Here $N_{1}$ denotes to the occupation of the ground level of PQW, $N_m$ is the occupation of the highest level, and $N_k$ is the occupation of $k$-level with $k=2\ldots m-1$, $\alpha$-terms describe the exciton-exciton scattering in the system. Terms $P_k^{(1)}(0)$ and $P_k^{(2)}(\tau_{delay})$ describe the initial two pulse excitation where the second pulse comes with a delay $\tau$.  Terms
$-N_k/\tau_k$ for $k=2\ldots m$ describe both radiative and non-radiative decay rate of excitons at each level. Phonon-assisted relaxation of excitons is taken into account in these terms. For the first level, only the radiative recombination is taken into account.

Matrix elements $W_{i}$ describe the transition from any level $k$ to any other level $k-i$ in the cascade.
Transitions between adjacent levels may be mediated by
emission of THz radiation as suggested in Ref.~\cite%
{liew_proposal_2013}. We generalize this model and consider the THz transitions
between all the levels, which is described in the above equations by summation over all levels. 

We assume that the pumping is centered at one of the middle levels of the
cascade, $k$,  meaning that upward scattering is possible from this level. The upward scattering due to exciton-exciton
interaction populates all the levels up to the highest one 
labeled $m$. For the structure under study, the exciton-exciton scattering is found to change the amplitude of the PL signal, but does not affect the most important features
of the exciton dynamics. The exciton-exciton scattering plays a minor role in our experiments and
the corresponding terms in rate equations can be safely omitted.

Experimentally, the system is excited by femtosecond pulses
which are relatively broad in energy and capable of pumping several energy
levels of the cascade simultaneously. To account for the spectral broadening
of the pulse in the model, we assume that polaritons are excited not only at
the level $k$, but also at the nearest levels $k-1$ and $k+1$. In the numerical simulations, the
cascade is considered to have maximum number  exciton levels $m=9$ with
level $k=6$ receiving the major part of input pulses power, $2P/3$, and levels $k=5$,~7, receiving 1/6 of total input power each.
We have used the following parameters in the calculations: $W_{1}=1500$~s$^{-1}$, $W_{2}=W_{4}=500$~s$^{-1}$, $W_{3}=100$~s$^{-1}$, $W_{5}=2500$~s$^{-1}$, $\tau _{g}=11$~ps, $\tau _{p}=55$~ps, $%
\tau _{u}=22$ ps. where $\tau _{g},\tau _{u},\tau _{p}$ are the decay
times for the ground level, pumped levels and levels above the pumped one,
respectively. 
\begin{figure}[h]
 \centering
 \includegraphics[width=1\linewidth]{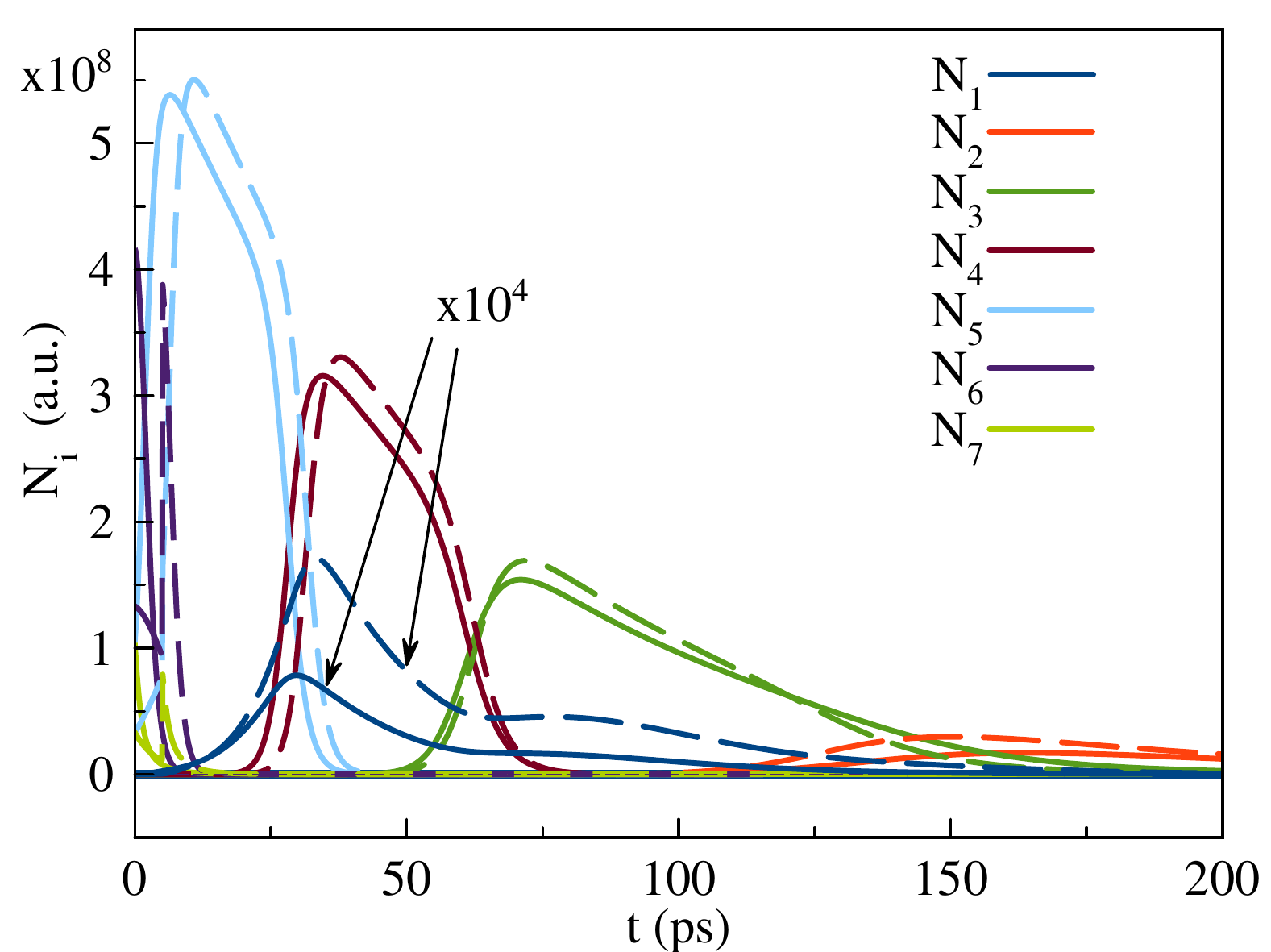}
\caption{(color online) Time evolution of exciton densities at each level in the QW. Solid lines are calculated for zero-delay between pulses and dashed lines show the same for $\tau_{delay}=-5$ps. Insets (a) and (b) show the same graph in logarithmic scale for zero and negative delays, respectively.} 
\label{fig:6}
\end{figure}

Fig.~\ref{fig:6} shows the time evolution of exciton densities at each energy level in the QW, plotted for two different delays between the pulses. As seen from the figure, the population dynamics is quite complex. If the system is excited by a single pulse (solid lines in Fig.~\ref{fig:6}), pumped levels 5 – 7 are populated and other levels are almost empty at the initial time interval ($t < 10$~ps) after the pulse. Due to the high exciton density at the level 5, $N_5 >> 1$, the Bose-stimulated relaxation from the upper levels $6$ and $7$ is switched on and the population of this level dramatically increases. The population of level 5 reaches its maximum at $t \approx 5$~ps while levels 6 and 7 become empty. The low-lying levels, $i = 4 \ldots 1$, are slowly populated while the exciton density is not reached a critical value for Bose-stimulated relaxation. This critical value is achieved for the level $4$ first because, in the framework of our model, the relaxation between adjacent levels is more efficient, $W_1 > W_2, \dots, W_4$. This explains the threshold-like increase of population of the level $4$ at time $t \approx 30$~ps. Similarly, populations of levels $3$ and $2$ is rapidly increase at time $t \approx 60$~ps and $t \approx 120$~ps, respectively [see respective curves in Fig.~\ref{fig:6}]. However, the population of the lowest exciton level, $1$, is not efficiently boosted via this pathway because of the low population of the adjacent level $2$. Therefore we have to assume in our model that there is a direct relaxation of excitons from the pumped level $5$ to the lowest level. As we will see below, this process explains the second maximum observed experimentally in pump-pump experiments [see Fig.~\ref{fig:5}~(a)].

\begin{figure}[h]

 \centering
 \includegraphics[scale=0.27]{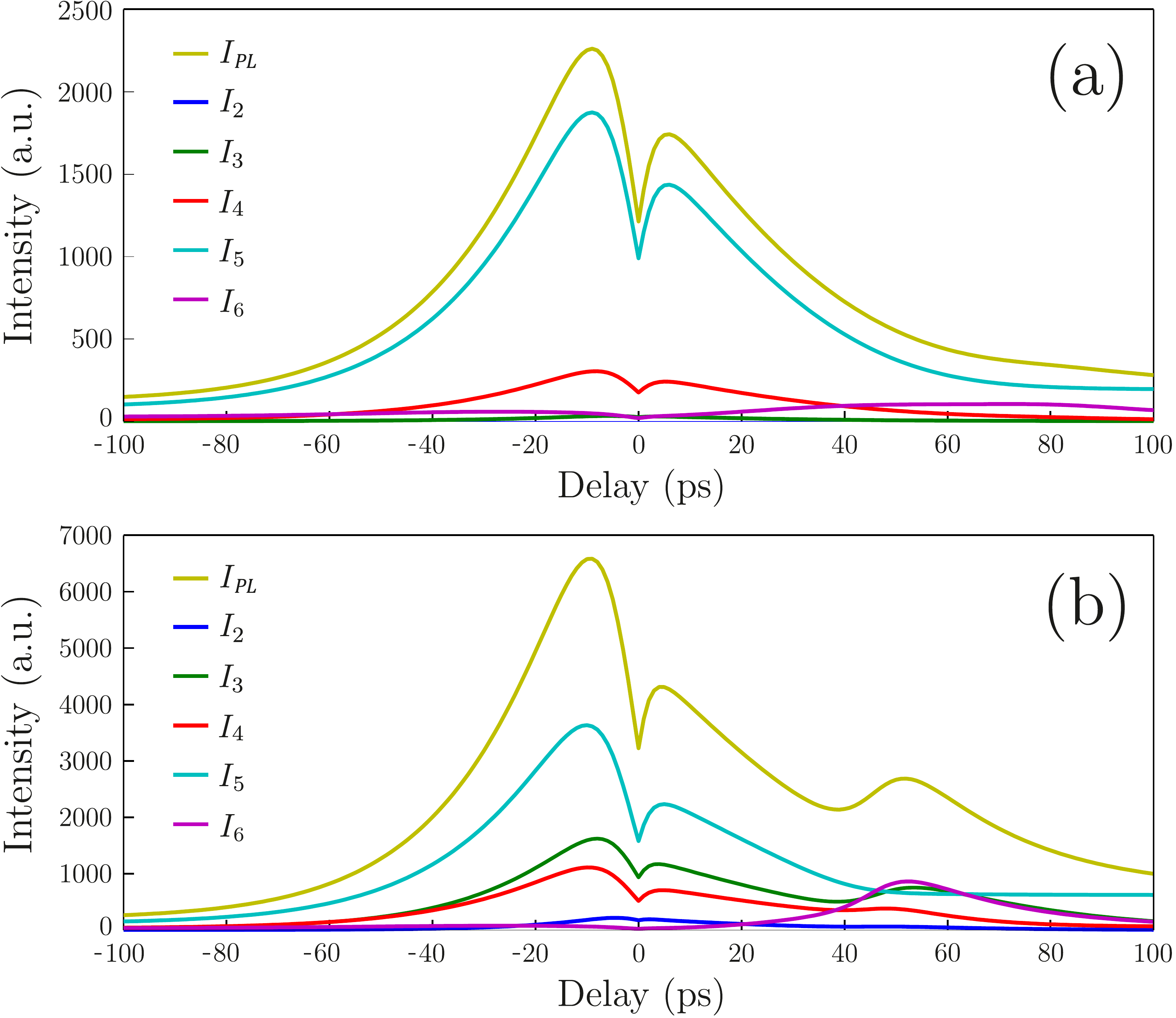}
\caption{(color online) Modeling of pump-pump signal with use of rate equations: delay dependencies of total PL intensity $I_{PL}$ from the ground exciton level, and separate contributions  of each transition $I_i$ =$W_{i-1}N_{i}(N_1+1)$ for $i=2..6$, plotted for two different values of pump powers: (a) $P^{(1)}$=0.3$\times P_{Thr}$, $P^{(2)}$%
=0.25$\times P_{Thr}$  and (b)$P^{(1)}$=0.7$\times P_{Thr}$, $P^{(2)}$=0.25$\times
P_{Thr}$. }
\label{fig:7}
\end{figure}

Nonradiative losses of excitons, described by terms $-N_i/\tau_j$ in EQs.~(3) – (5), compete with the relaxation processes. The integral magnitude of losses depends on the time, spent by excitons are at the excited levels. This time can be drastically shortened and, correspondingly, the PL yield can be increased, if appropriate experimental conditions initiating the Bose-stimulated relaxation are fullfiled. In particular, the excitation power, which should be close to the threshold power for polariton lasing~\cite{kavokin_cavity_????-1}. Separation of the excitation pulse in two pulses also helps controlling the population of different exciton levels (see Fig.~\ref{fig:6}) and, hence, the nonradiative losses. Once the excitons created by the first pulse have relaxed to the 5th level (it takes of about 5~ps), the excitons created by the second pulse delayed by $\tau = 5$~ps rapidly relax from the 6th and 7th levels to the 5th one via Bose-stimulated process. This stimulation gives rise to the increased population of level 5 relative to that obtained for zero delay, as one can conclude comparing solid and dashed lines $N_5$ in Fig.~\ref{fig:6}. The corresponding increase of population is observed also for other levels. In particular, a remarkable increase of population is observed for level 1, which is the key point for understanding of the dip in the delay dependence of PL intensity observed experimentally, see Fig.~\ref{fig:5}~(a). 

Figure~\ref{fig:7} shows the integral PL intensity, $I_{PL}$, as a function of the delay between pulses, $\tau_{delay}$, for two excitation powers with total power, $P^{(1)} + P^{(2)} < P_{th}$, where $P_{th}$ is the threshold power for polariton lasing. Curves $I_i$ represent the contribution of each transition term having form $W_{i-1} N_{i}(N_1 +1)$ for $i = 2 \ldots 6$, into the total PL. As one can  see from the figure, the modelling predicts a dip in the PL intensity at the small delays. It is clear from the discussion above that the dip is due to the increase of PL intensity at the delay increases up to several picoseconds.

Further increase of delay between the pulses gives rise to the depopulation of level 5 when the second pulse arrives. As a result the Bose-stimulated relaxation of levels 6 and 7 excited by the second pulse becomes less efficient and the nonradiative losses increase. This explains the decrease of PL intensity at delays $\tau = 10 \ldots 40$~ps, see Fig.~\ref{fig:7}~(b). However, when the delay $\tau > 40$~ps, the population of the ground exciton level is so large ($N_1 >> 1$) that the direct Bose-stimulated relaxation from level 6 described by term $W_5 N_6 (N_1 +1)$ becomes efficient pathway for the exciton relaxation to the ground level. Correspondingly, efficient depopulation of level 6 occurs that results in the decrease of nonradiative losses. These processes explain the appearance of second peak at the delay dependence of PL intensity. The calculated behavior of total PL intensities at weak and strong pumping qualitatively reproduce the experimental
results (compare with Fig~\ref{fig:5}~(a))

\section{Conclusions}

Present experiments and modelling shed light on the exciton dynamics in bosonic cascades; the pump-pump method being a powerful tool for the study of the
%femtosecond-scale 
fast relaxation dynamics at the non-resonant pumping. 
When the only one pump pulse is used for excitation, the relaxation occurs via one pathway. 
Using the second pulse allows one to switch the relaxation between different pathways depending on delay between the pulses which we demonstrate experimentally and through modelling.  
We have found a qualitative agreement between the theoretical model of a BCL in a PQW system and experimental results.
Because the pump-pump method is based on strong non-linearity of PL yield on the pump power, which is close to the threshold, we could not expect the quantitative agreement of the theory and the experiment. However, the modelling showed that there are two different pathways for relaxation in the system. The first pathway is the relaxation via cascade transitions, where all levels are being filled, and the second pathway is the direct transition from the pumped level to the ground one. By taking these pathways into account, the model may be generalized for larger number of levels or for other initial conditions. Taking into account the fact that minimum in the PL at zero delay occurs only if polaritons are exited on at least two adjacent levels, it is possible to explain the difference in PL for PQW and bare QW shown in Fig~\ref{fig:5}: the levels in bare QW stand far from each other and polaritons are excited only at one energy level. Due to this there is no minimum at zero delay between pump pulses. However the relaxation process in the bare QW still may be described by the rate equations, but with different parameters values.

As a conclusion this
work shows potentiality of microcavities with embedded PQW for realization
of Bosonic Cascade Lasers.

\begin{acknowledgments}

Acknowledgments:
The authors thank Dr. Anton Nalitov for fruitful discussions.
A.K. and E.C. thank the EPSRC established career fellowship and EPSRC Hybrid Polaritonic Programme, A.K. is grateful for the
financial support from the Russian Ministry of Science and Education
(contract no. 11.G34.31.0067 and  RFBR Grants No.  15-59-30406. I.I. thanks financial support of SPbU (grants No. 11.38.213.2014).   
 The authors also thank the SPbU Resource Center ``Nanophotonics'' 
(www.photon.spbu.ru) for the sample studied in present work. 

\end{acknowledgments}

\bibliography{bibliography2}

\end{document}